\documentclass{article}
\usepackage{makecell}
\usepackage{spconf,amsmath,graphicx,hyperref}

\usepackage{xcolor}
\usepackage{booktabs}
\usepackage{multirow} 
\usepackage{bm}
\usepackage{amsmath,amssymb,mathtools}
\usepackage{graphicx}
\usepackage{xcolor}
\usepackage{booktabs}   %
\usepackage{hyperref}   %
\usepackage{algorithm}
\usepackage{algpseudocode}

\usepackage{xcolor}
\usepackage[utf8]{inputenc} 
\usepackage[T1]{fontenc}
\usepackage{url}
\usepackage{ifthen}
\usepackage{cite}
\usepackage{subfig}

\hypersetup{colorlinks,
	linkcolor=blue,%
	citecolor=blue}
    
\usepackage{hyperref}

\usepackage{caption}
\usepackage{comment}
\usepackage{graphicx}   
\usepackage{pifont}     

\title{Leveraging Overfitting for Low-Complexity and \\
Modality-Agnostic Joint Source-Channel Coding}

\name{Haotian Wu, Gen Li, Pier Luigi Dragotti, Deniz Gündüz{\thanks{Contact: haotian.wu17@imperial.ac.uk. \newline\indent\quad Project page: {https://eedavidwu.github.io/Implicit-JSCC/}}}}

\address{Department of Electrical and Electronic Engineering, Imperial College London, UK}

\begin{document}

\maketitle

\begin{abstract}
This paper introduces Implicit-JSCC, a novel overfitted joint source–channel coding paradigm that directly optimizes channel symbols and a lightweight neural decoder for each source. This instance-specific strategy eliminates the need for training datasets or pre-trained models, enabling a storage-free, modality-agnostic solution. As a low-complexity alternative, Implicit-JSCC achieves efficient image transmission with around $1000\times$ lower decoding complexity, using as few as $607$ model parameters and $641$ multiplications per pixel. This overfitted design inherently addresses source generalizability and achieves state-of-the-art results in the high SNR regimes, underscoring its promise for future communication systems, especially streaming scenarios where one-time offline encoding supports multiple online decoding.
\end{abstract}
\section{Introduction}
Advances in wireless communication and machine learning have enabled edge intelligence, but growing data and latency demands challenge resource-limited devices. While Shannon’s separation theorem holds in the infinite block length regime, it breaks down under practical constraints, motivating joint source-channel coding (JSCC). Although classical JSCC remains difficult to design, recent deep learning advances have led to competitive DeepJSCC schemes \cite{bourtsoulatze2019deep,wu2025deep,9791398,xu2021wireless,wu2023transformer,yang2024swinjscc,10480348}.

The first DeepJSCC scheme for image transmission was proposed in \cite{bourtsoulatze2019deep}, outperforming conventional approaches that combine better portable graphics (BPG) with low-density parity-check (LDPC) codes. This approach was later extended to various scenarios, such as multiple-input multiple-output (MIMO) \cite{wu2024deep} and relay channels \cite{bian2025process}. With task-specific training, DeepJSCC can also be applied to various modalities, showing strong adaptability and efficiency \cite{wu2025deep}. These advantages position DeepJSCC as a promising foundation for emerging semantic communication systems \cite{Gunduz:JSAC:23}.

\begin{figure}[t] 
    \centerline{\includegraphics[width=0.98\linewidth,trim=1cm 2.65cm 11.2cm 2.1cm, clip]{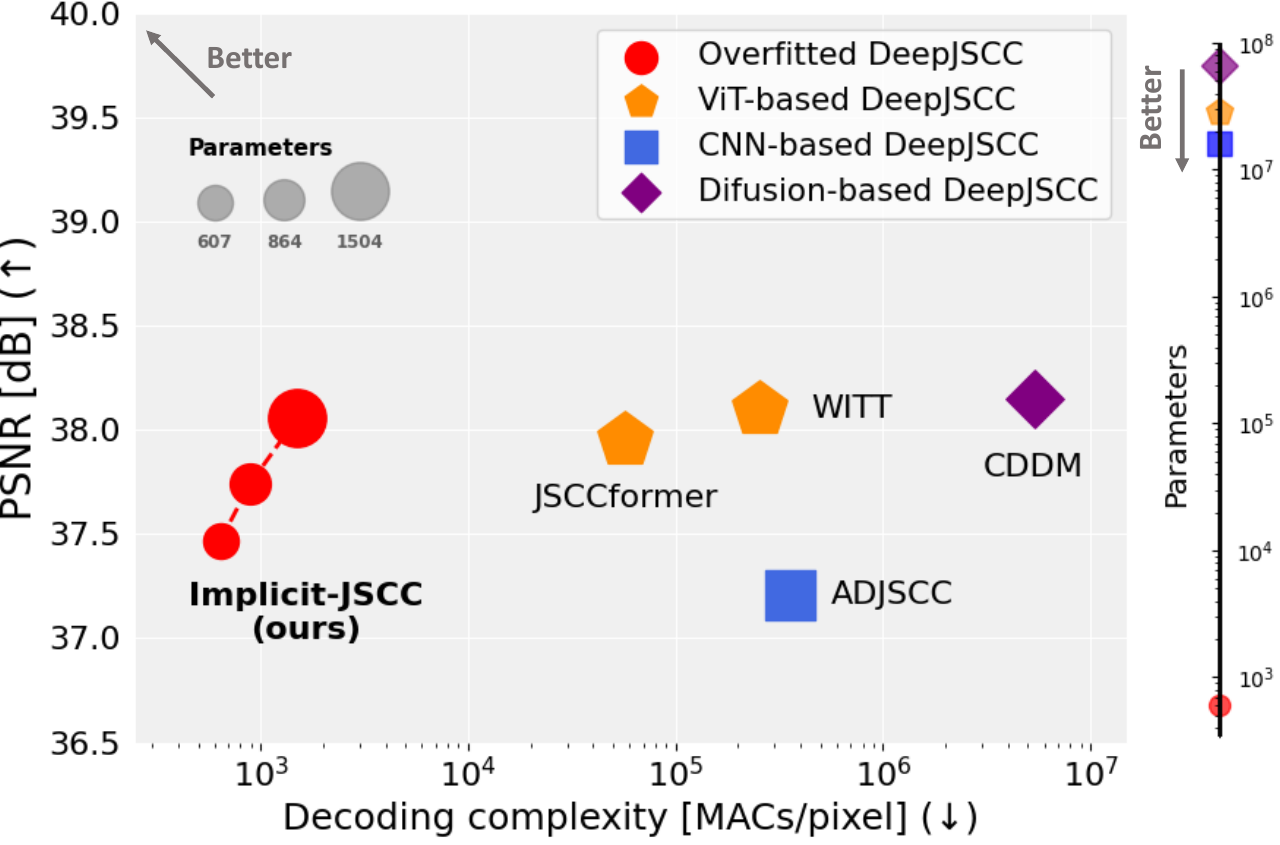}}
    \caption{Performance vs. decoding complexity on Kodak (SNR=$10$ dB and R=$0.229$), where Implicit-JSCC achieves strong performance with three orders of magnitude fewer decoding operations and significantly fewer parameters.}
    \label{psnr_complexity_kodak}
\end{figure}

However, the exponential growth of multimodal and compute-intensive edge tasks exposes new challenges in current DeepJSCC methods, underscoring the need for more flexible, source-adaptive, and efficient designs. Specifically, current approaches still suffer from the following limitations: 

\textbf{Decoding complexity:} Competitive DeepJSCC schemes employ advanced architectures\cite{bourtsoulatze2019deep,wu2025deep,9791398,xu2021wireless,wu2023transformer,yang2024swinjscc,10480348,xu2025semantic}, which incur high decoding costs (Fig.~\ref{psnr_complexity_kodak}), limiting their use on edge devices.

\textbf{Source-generalizability:} Current DeepJSCC methods \cite{bourtsoulatze2019deep,xu2021wireless,yang2024swinjscc} rely on large training datasets and exhibit degraded performance under distribution shifts~\cite{kulinski2023towards}, often requiring costly retraining or new data collection.

\textbf{Modality-generalizability:} Most existing DeepJSCC schemes are modality-specific \cite{bourtsoulatze2019deep,farsad2018deep,Tung:JSAC22}, requiring separate models and mechanisms for each modality, with task-dependent model switching at the transceiver.    

\textbf{Model storage overhead:} Most DeepJSCC schemes rely on autoencoders \cite{bourtsoulatze2019deep,wu2025deep,9791398,xu2021wireless,wu2023transformer,yang2024swinjscc}, requiring large model storage at both ends. Supporting diverse scenarios demands separate models, further increasing memory usage and limiting scalability.

Inspired by recent advances in overfitted source coding \cite{dupont2022coin++,wu2024mimo,ladune2023cool,kim2024c3,LotteryCodec}, we propose Implicit-JSCC, a novel overfitted DeepJSCC paradigm that encodes each source instance into channel symbols and a lightweight neural decoder. As illustrated in Fig.~\ref{fig_1}, this instance-specific design eliminates the need for training data, drastically reduces decoding complexity, and achieves strong performance. By transmitting the decoder directly, Implicit-JSCC provides a memory-efficient, modality-agnostic solution that avoids model switching and enables flexible deployment. Our main contributions are:


\begin{itemize}
\item We present Implicit-JSCC, the first overfitted DeepJSCC that encodes each source instance directly into channel symbols, providing a low-complexity decoding and a new design perspective.

\item Leveraging instance-specific optimization, Implicit-JSCC eliminates the need for training sets and effectively addresses source and modality generalization.

\item Implicit-JSCC reduces decoding complexity by up to $1000$× over standard schemes, offering a low-complexity alternative without pre-trained storage for efficient and flexible deployment on edge devices.

\item Extensive experiments show that Implicit-JSCC delivers strong performance across diverse scenarios, highlighting the potential of instance-specific optimization in advancing DeepJSCC.

\end{itemize}
\section{System model}
We consider transmitting a source sample $\bm{S}\in \mathbb{R}^{N\times C}$ over $k$ uses of a wireless channel, where $C$ is the number of channels and $N$ denotes the number of source symbols per channel. This corresponds to a \textit{bandwidth ratio} of $R\triangleq\frac{k}{NC}$. 
\subsection{Problem formulation} 
\noindent\textbf{Transmitter.} An encoder $\mathcal{E}(\cdot): \mathbb{R}^{N\times C} \rightarrow \mathbb{C}^k$ maps the source $\bm{S}$ into channel symbols $\bm{X}\in \mathbb{C}^k$ as:
\begin{equation}
\bm{X}=\mathcal{E}(\bm{S}),
\end{equation}
where $\bm{X}$ is subject to a power constraint: $\frac{1}{k} \mathbb{E}\left[\|\bm{X}\|^2_\text{2}\right]\leq 1$.

\noindent\textbf{Channel model.} Subsequently, $\bm{X}$ is transmitted over an additive white Gaussian noise (AWGN) channel, modeled as: $\bm{{Y}}=\mathcal{H}(\bm{X})= \bm{X}+\bm{W}$, where $\bm{{Y}}\in \mathbb{C}^{k}$ is the channel output and $\bm{W}\in \mathbb{C}^{k}$ denotes the AWGN term, with each element $W[i] \sim \mathcal{CN}(0,\sigma_w^2)$. The signal-to-noise ratio (SNR) is defined as $\mu \triangleq 10\log_{10} \left(\frac{1}{\sigma_w^2}\right) \text{ dB}$ and is known at both ends.
\begin{figure}[t] 
    \centering
\includegraphics[width=0.9\columnwidth]{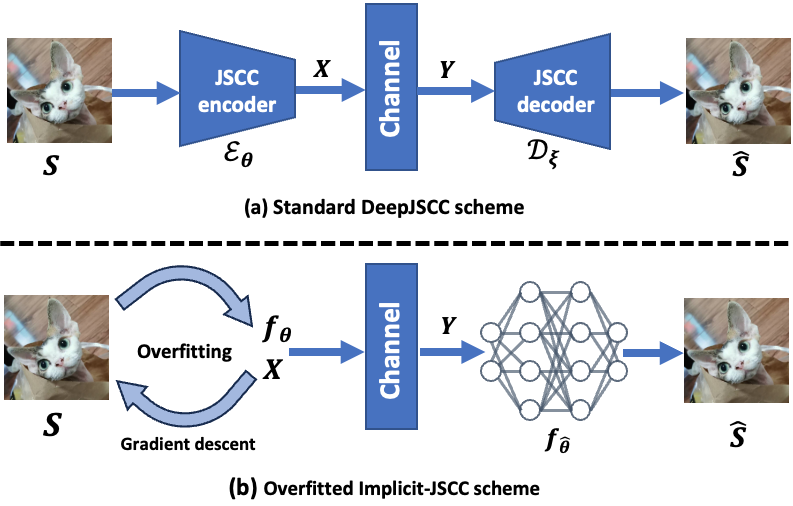}
    \caption{Illustration of alternative DeepJSCC schemes. Top: Standard DeepJSCC with pre-trained auto-encoders. Bottom: Overfitted DeepJSCC, where channel symbols and a lightweight decoding function are optimized per sample.}
    \label{fig_1}
\end{figure}

\noindent\textbf{Receiver.} A decoder $\mathcal{D}(\cdot):  \mathbb{C}^k\rightarrow \mathbb{R}^{N\times C}$ reconstructs the source from ${\bm{Y}}$ as:
\begin{equation}
   \bm{\hat{S}}=\mathcal{D}(\bm{Y}),
\end{equation}
where $\bm{\hat{S}}\in \mathbb{R}^{N\times C}$ represents the reconstructed source. 

\subsection{Standard DeepJSCC}
Standard DeepJSCC \cite{bourtsoulatze2019deep} jointly trains an encoder–decoder pair, $\mathcal{E}_{\bm{\theta}}(\cdot)$ and $\mathcal{D}_{\bm{\phi}}(\cdot)$, on a large training dataset:
$(\bm{\theta}^*,\bm{\phi}^* )=\mathop{\arg\min}_{\bm{\theta},\bm{\phi}}\mathbb{E}\big[\mathcal{L}(\bm{S},\bm{\hat{S}})\big]$, where $\mathcal{L}(\cdot)$ denotes the distortion loss. Once well-trained, the transmitter generates the channel symbols $\bm{X^{i}}$ \footnote{\small{In this paper, the notation $\bm{S}^i$ with a superscript denotes a specific instance of the random variable $\bm{S}$.}}, and the receiver reconstructs $\bm{\hat{S}^{i}}$ as:
\begin{equation}
    \bm{X^{i}}=\mathcal{E}_{\bm{\theta^*}}(\bm{S^{i}}), \quad \bm{\hat{S}^{i}}=\mathcal{D}_{\bm{\phi^*}}(\bm{Y^{i}}).
\end{equation}
\subsection{Overfitted DeepJSCC}
We propose an alternative strategy: overfitting a distinct codec for each instance, termed {overfitted DeepJSCC} (see Fig. \ref{fig_1}). For each $\bm{S^{i}}$, the transmitter jointly optimizes the channel symbols $\bm{X^{i}}$ and a lightweight neural function $f_{\bm{\theta^i}}$: 
\begin{equation}
( \bm{X^{i}},\bm{\theta^{i}}) = \mathcal{O}(\bm{S^{i}}),
\end{equation}
where $\mathcal{O}(\cdot)$ denotes the overfitting process. The pair $\{\bm{X}^i, \bm{\theta}^i\}$ forms a sample-specific codec\footnote{\small{Neural parameters $\bm{\theta^i}$ may serve as the sole codeword (i.e., $\bm{X^i}\triangleq \bm{\theta^i}$), turning source transmission into function delivery.}}. At the receiver, the retrieved function $\bm{\hat{\theta}^i}$ reconstructs the source: $\bm{\hat{S}^{i}}=f_{\bm{\hat{\theta}^i}}(\bm{Y^{i}})$. 

\section{Implicit-JSCC}
This section presents Implicit-JSCC, a practical overfitted DeepJSCC method, which overfits each source instance into channel inputs and a compact neural function.
\subsection{{Channel input construction}} 
\begin{figure}[t]
\centering
\includegraphics[width=0.975\columnwidth]{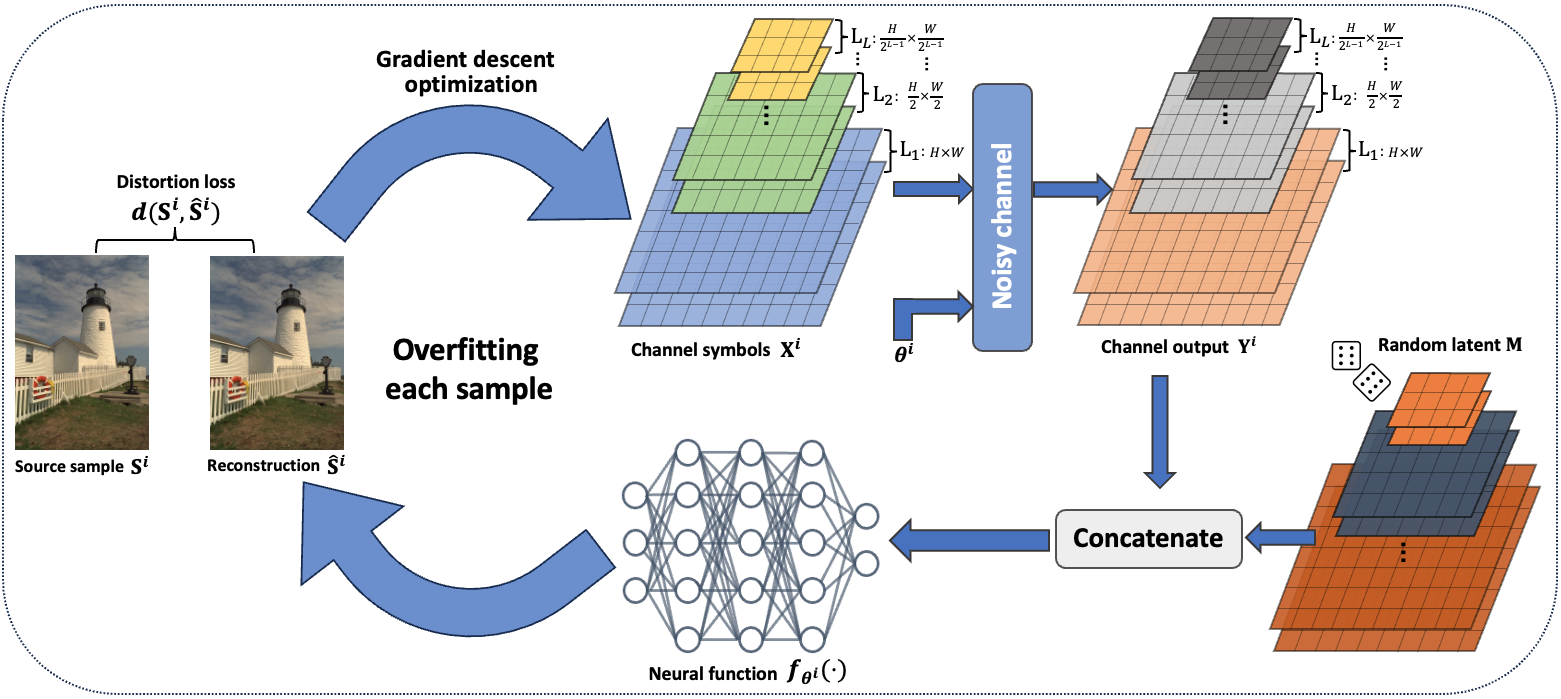}
\caption{Implicit-JSCC transmitter: jointly optimizes neural function and channel symbols via gradient descent algorithm.}
\label{OF_Transmitter}
\end{figure}

As shown in Fig. \ref{OF_Transmitter}, for each sample, channel input $\bm{X^{i}}$ is constructed using an $L$-resolution pyramid structure: 
\begin{equation}
\bm{X^{i}} \triangleq \{{\bm{{x}_1^i},\bm{{x}_2^i},\ldots,\bm{{x}_{L}^i}}\},
\end{equation}
where each $\bm{{x}_k^i}\in \mathbb{R}^{L_k\times \frac{N}{4^{k-1}}}$ is a learnable matrix, and $\bm{X^{i}}$ is reshaped into $\mathbb{C}^{\sum\limits_{k=1}^{L} {\frac{L_k N}{2\cdot 4^{k-1}}}}$ for transmission. The corresponding channel output, after being reshaped into the real domain, is: 
$\bm{Y^{i}} \triangleq \{{\bm{{y}_1^i},\ldots,\bm{{y}_{L}^i}}\}$, 
where $\bm{y_k^i}=\mathcal{H}(\bm{x_k^i})\in \mathbb{R}^{L_k\times \frac{N}{4^{k-1}}}$. The resultant bandwidth ratio is $R_x=\sum\limits_{k=1}^{L} {\frac{L_k}{2C\cdot 4^{k-1}}}$, where $L_k$ and $L$ can be adjusted depending on available bandwidth.

\subsection{{Neural function design}} 
\begin{figure}[t]
\centering
\includegraphics[width=1\columnwidth]{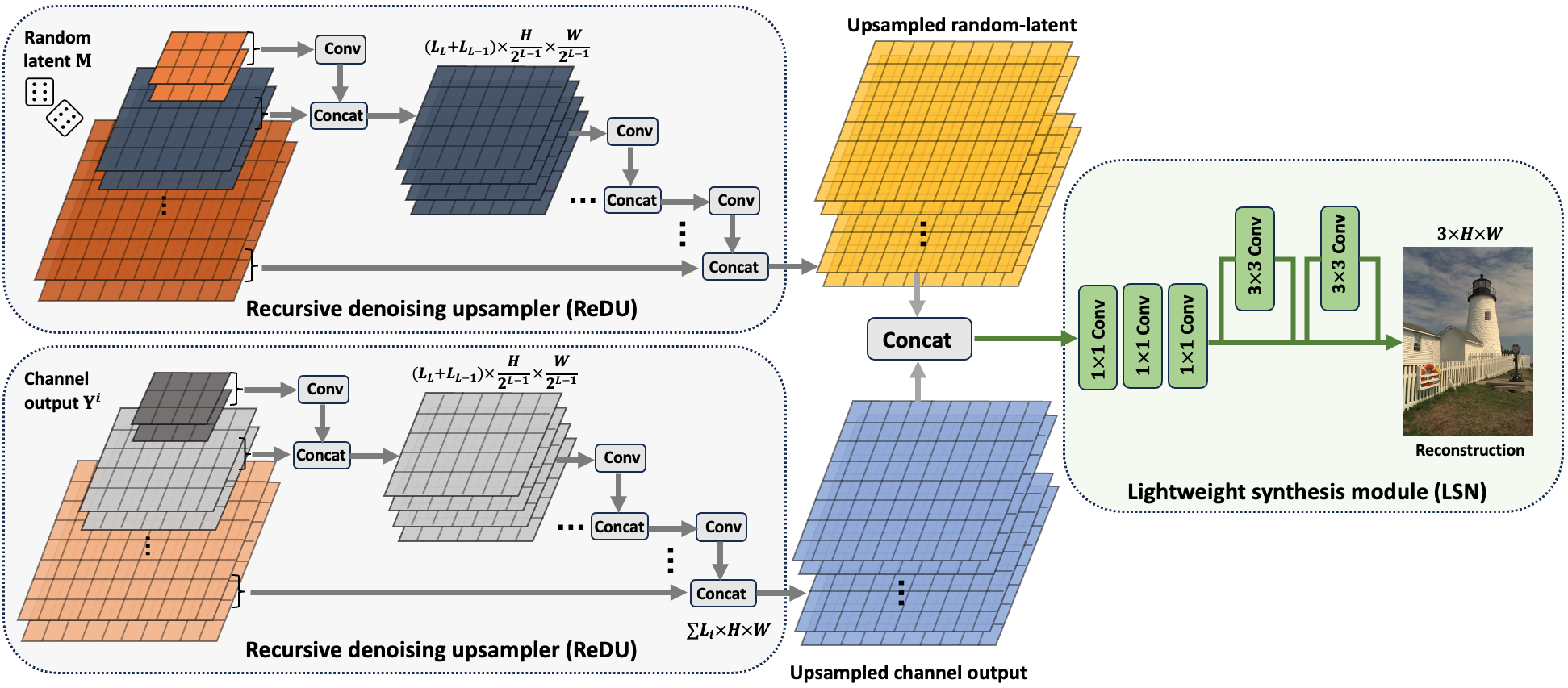}
\caption{Implicit-JSCC receiver: leverages common randomness to integrate channel condition and enhance performance.}
\label{OF_receiver}
\end{figure}

As illustrated in Fig. \ref{OF_receiver}, the neural function $f_{\bm{\theta}}$ comprises a Recursive Denoising Upsampler (ReDU) and a Lightweight Synthesis Module (LSM). 

\textbf{ReDU module} is a transposed convolutional kernel with a kernel size of $8$, serving as a denoiser and refiner. Formally, it performs recursive upsampling starting from the lowest-resolution channel output. Given $\bm{y_L^i}\in \mathbb{R}^{L_L\times \frac{N}{4^{L-1}}}$, the process begins with $\bm{U_{L}^i}\triangleq\textit{Upsample}(\bm{y_L^i})\in \mathbb{R}^{L_L\times \frac{N}{4^{L-2}}}$, and continues recursively as:
\begin{align}
    \bm{u_{k-1}^i} &= \textit{Concatenate}(\bm{y_{k-1}^i}, \bm{U_{k}^i}), \quad k \in \{L,\dots,2\},\\
    \bm{U_{k-1}^i} &= \textit{Upsample}(\bm{u_{k-1}^i}), \quad k \in \{L,L-1,\dots,3\}.
\end{align}
Here, $\bm{U_{k-1}^i}$ represents the upsampling result at each iteration, while $\bm{u_{k-1}^i}$ denotes the intermediate concatenated result. 
This recursive process continues, concatenating and upsampling up to $\bm{y_1^i}$, producing 
$\bm{{U}^{i}}\triangleq \bm{u_1^i}\in \mathbb{R}^{\big(\sum\limits_{k=1}^{L}L_k\big)\times {N}}$.

\textbf{LSM module} is cascaded with ReDU and serves as a synthesis module to reconstruct $\bm{S^i}$ from $\bm{U^i}$. To incorporate channel information and improve performance, LSM introduces common randomness via a random matrix $\bm{M} \sim \mathcal{N}(0, \sigma_w^2)$, matching the shape of $\bm{Y^i}$. Using a synchronized seed, $\bm{M}$ is generated at both ends without transmission cost. After ReDU, $\bm{M}$ is concatenated with $\bm{U^i}$ and fed to LSM for the final reconstruction. Specifically, LSM comprises three $1 \times 1$ convolutions with weights $\bm{W_1^i} \in \mathbb{R}^{\big(2\sum_{k=1}^{L}L_k\big)\times d}$, $\bm{W_2^i} \in \mathbb{R}^{d \times d}$, and $\bm{W_3^i} \in \mathbb{R}^{d \times C}$, followed by two $3 \times 3$ convolutions outputting $C$ channels, resulting in the final reconstruction $\bm{\hat{S}^i} \in \mathbb{R}^{N \times C}$.
\begin{figure*}[tb]
    \centering
    \subfloat[]{
    \begin{minipage}[t]{0.23\linewidth}
    \centering
\centerline{\includegraphics[width=\linewidth,trim=1cm 0.2cm 1cm 1.5cm, clip]{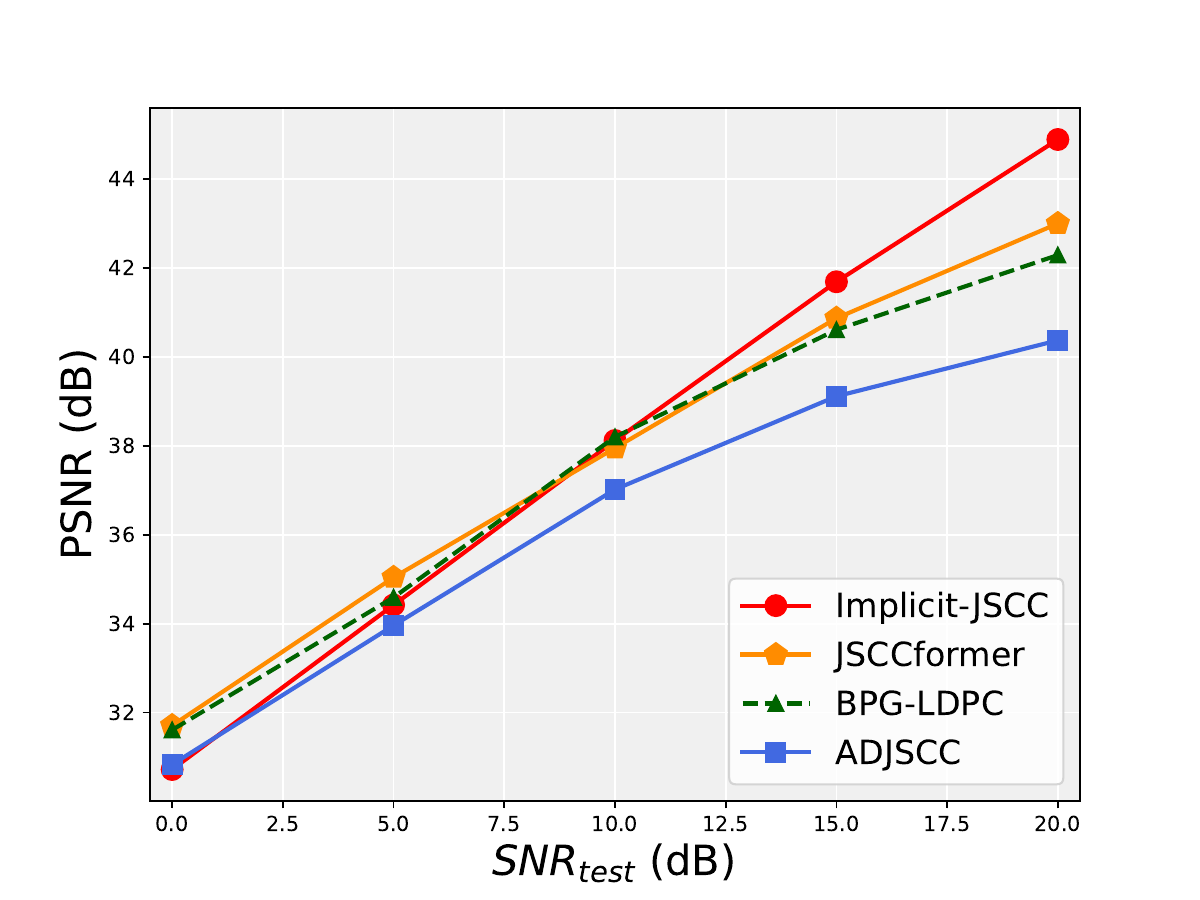}}
    \label{JSCC_KODAK_PSNR}
    \end{minipage}%
    }%
     \hfill
    \subfloat[]{
    \begin{minipage}[t]{0.23\linewidth}
    \centering
\centerline{\includegraphics[width=\linewidth,trim=1cm 0.2cm 1cm 1.5cm, clip]{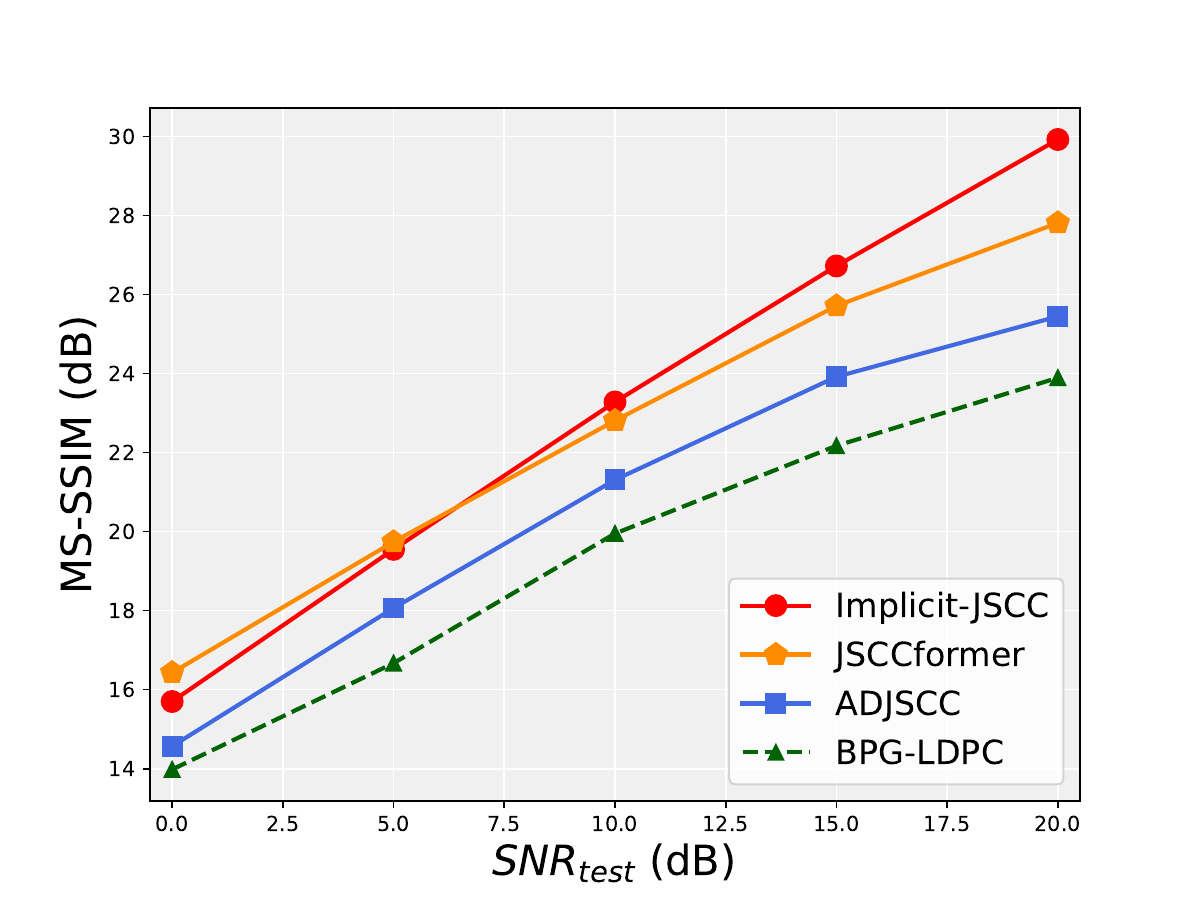}}
    \label{JSCC_KODAK_SSIM}
    \end{minipage}%
    }%
     \hfill
    \subfloat[]{
    \begin{minipage}[t]{0.24\linewidth}
    \centering
    \centerline{\includegraphics[width=\linewidth,trim=2.25cm 1.8cm 3cm 1.8cm, clip]{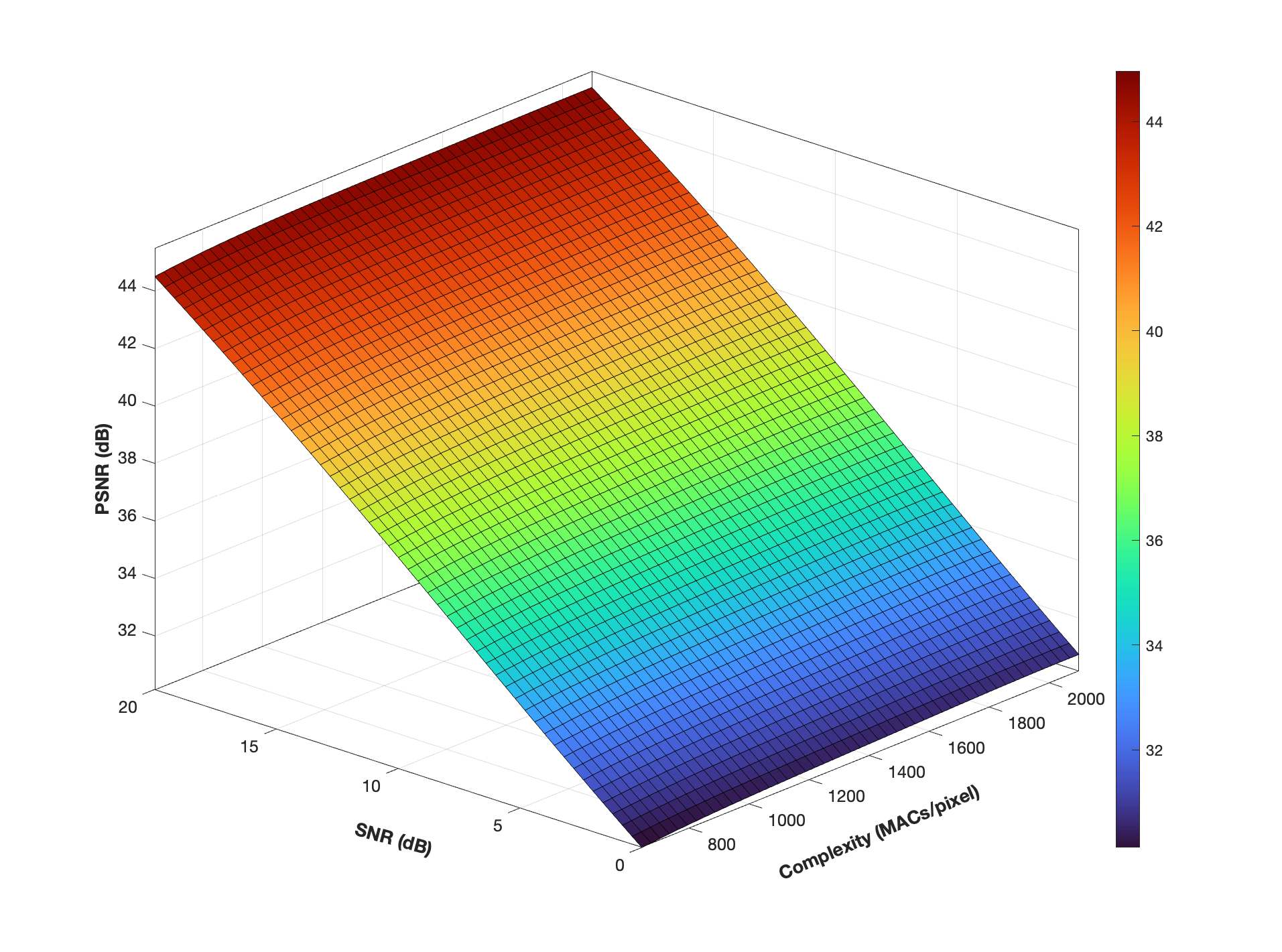}}
        \label{JSCC_KODAK_PSNR_complexity_SNR}
    \end{minipage}%
    }%
     \hfill
    \subfloat[]{
    \begin{minipage}[t]{0.23\linewidth}
    \centering
    \centerline{\includegraphics[width=\linewidth,trim=1cm 0.2cm 1cm 1.5cm,clip]{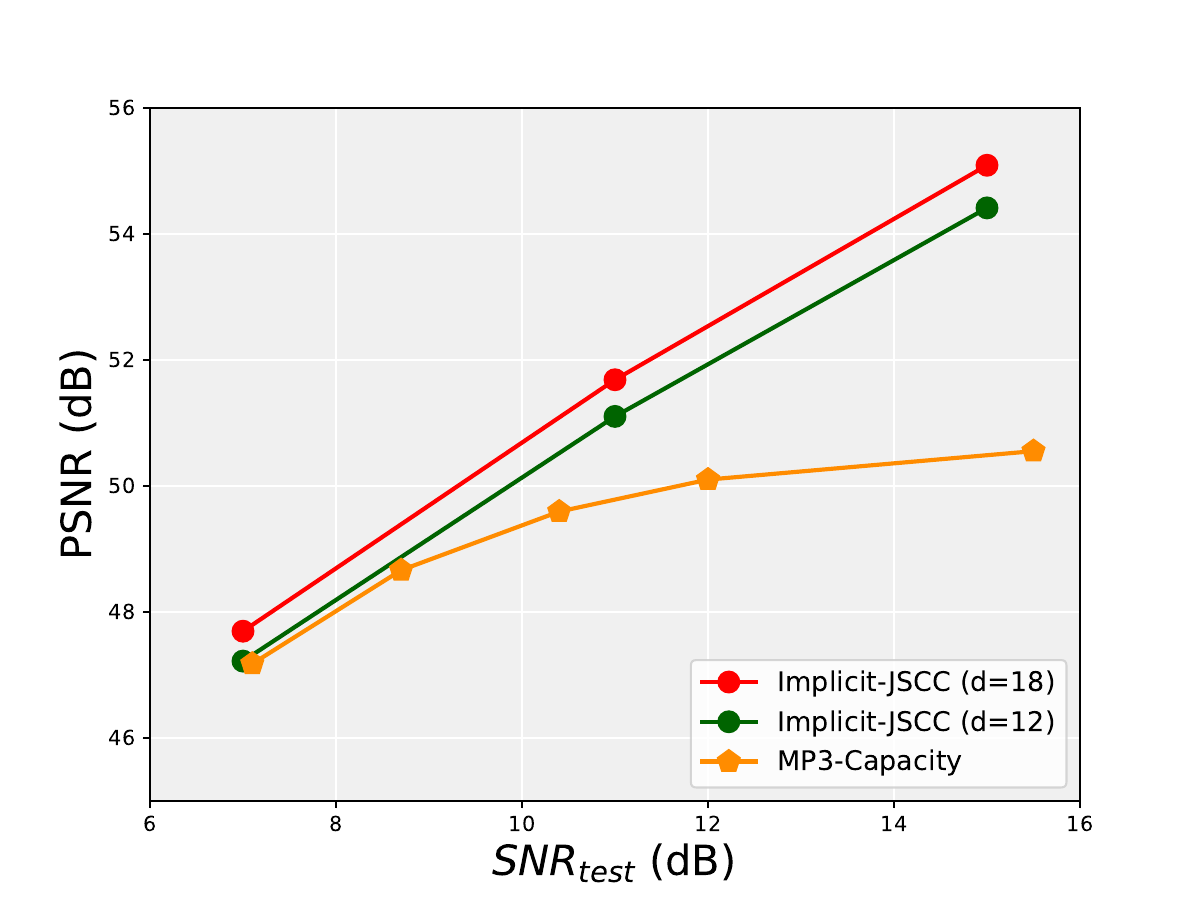}}
    \label{JSCC_mp3}
    \end{minipage}%
    }%
    \centering
    \caption{Performance of Implicit-JSCC. (a)–(c): Image transmission results on Kodak showing PSNR, MS-SSIM, and the trade-off between complexity and distortion over varying SNRs. (d): Audio transmission performance across SNRs.}
    \label{JSCC_performance} 
\end{figure*}
\begin{figure*}[t]
\centering
\begin{minipage}[t]{0.58\textwidth}
\vspace{0pt}
  \centering
  \centerline{\includegraphics[width=1\linewidth,trim=1cm 6cm 4.5cm 0.2cm, clip]{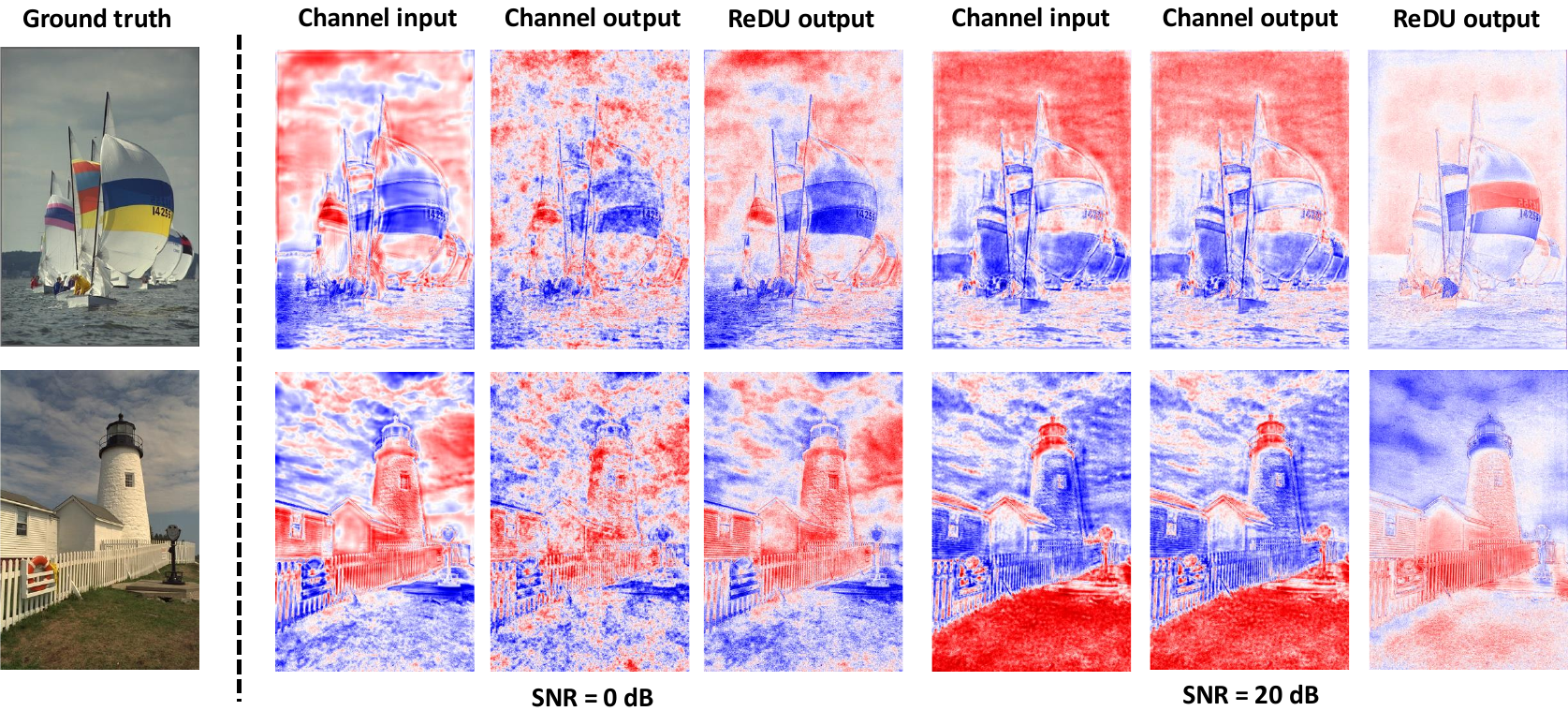}}
  \captionof{figure}{Visualization of Implicit-JSCC channel symbols and ReDU output.}
  \label{fig_Vis_each_module}
\end{minipage}%
\hfill
\begin{minipage}[t]{0.4\textwidth}
  \centering
  \vspace{0pt}
  \captionof{table}{PSNR and coding time for different number of encoding steps on the first $10$ images of the Kodak dataset across different SNRs.}
  \label{tab_Encoding_complexity}
  \small        
  \setlength{\tabcolsep}{4pt}  
      \begin{tabular}{cccc}
   {\textbf{Encoding} }& \multirow{1}{*}{\textbf{PSNR for SNR}} &\textbf{Enc. time} & \textbf{Dec. time} \\
  \textbf{steps}  & $0$ dB / $10$ dB& \textbf{(s)} & \textbf{(ms)} \\
    \hline
     1,000 & ${24.48}$ / ${28.17}$& $15.07$ &  $2.08\pm0.11$ \\
    2,000 & ${26.54}$ / ${31.31}$& $30.04$ & $2.08\pm0.11$ \\
    5,000& ${28.90}$ / ${34.76}$& $75.06$& $2.08\pm0.11$ \\
    $10,000$& ${29.78}$ / ${36.50}$& $150.01$ & $2.08\pm0.11$\\
     20,000& ${30.06}$ / ${37.26}$& $310.52$  & $2.08\pm0.11$\\
    $50,000$& ${30.27}$ / ${37.64}$& $740.94$  & $2.08\pm0.11$\\
     100,000& ${30.55}$ / ${38.12}$& $1498.68$  &  $2.08\pm0.11$\\
    \end{tabular}
\end{minipage}
\end{figure*}

\subsection{Parameter transmission}
Given the low cost of neural functions, we transmit model parameters directly using repetition coding. 
Let $\kappa_l$ and $\kappa_r$ denote the repetition factors for the LSN and ReDU, resulting a model bandwidth cost of $R_\theta = \frac{\kappa_l|\bm{\theta^i_{LSN}}|+\kappa_r|\bm{\theta^i_{ReDU}}|}{2NC}$, where $|\bm{\theta}^i_{\text{LSN}}|$ and $|\bm{\theta}^i_{\text{ReDU}}|$ denote their respective parameter counts. The total bandwidth ratio is $R=R_\theta+R_x$. To enable flexible coding across samples and channel conditions, we adapt $\kappa_l$, $\kappa_r$, and the hidden dimension $d$ via greedy search for best performance under fixed bandwidth or complexity budgets.


\color{black}
\subsection{Optimization strategy}
Implicit-JSCC employs a gradient descent approach to jointly optimize $\bm{X^{i}}$ and $\bm{\theta^{i}}$ by minimizing the distortion loss:
\begin{equation}
\bm{X^{i}}\leftarrow\bm{X^{i}}-\alpha\nabla_{\bm{X^{i}}}\mathcal{L}(f_{\bm{\hat{\theta}^{i}}}(\bm{Y^{i}}),\bm{S^{i}}),
\label{loss_1}
\end{equation}
\begin{equation}
\bm{\theta^{i}}\leftarrow\bm{\theta^{i}}-\alpha\nabla_{\bm{\theta^{i}}}\mathcal{L}(f_{\bm{\hat{\theta}^{i}}}(\bm{Y^{i})},\bm{S^{i}}),
\label{loss_2}
\end{equation}
where $\alpha$ is the learning rate and $\bm{\hat{\theta}^i}$ is the noisy parameters.
\color{black}
\section{Training and evaluation}
We evaluate Implicit-JSCC on image and audio transmission. For images, we compare against: ADJSCC \cite{xu2021wireless}, JSCCformer \cite{wu2023transformer}, WITT \cite{yang2024swinjscc}, CDDM \cite{10480348}, and BPG-LDPC (as configured in \cite{wu2024deep}), all trained on ImageNet and tested on Kodak using PSNR and MS-SSIM. For audio, we use the first 2-second clips of $30$ recordings from LibriSpeech `test-clean' split (as in \cite{dupont2022coin++}). MP3 with capacity-achieving channel code serves as a benchmark. Implicit-JSCC is trained for 100K steps with $\alpha=0.01$ and varying $(d, \kappa_l)$ pairs \footnote{\small{We employ five $(d, \kappa_l)$ pairs ($\{(12,25), (18,17), (24,12), (30$\newline$, 9), (36,7)\}$), selected via greedy search for best performance (Section \ref{section_performance}) and for varying complexity budgets (Fig. \ref{psnr_complexity_kodak}).}}. We consider a total bandwidth ratio of $R=0.2292$ (with $L=7$, $L_1=\cdots=L_7=1$, and $\kappa_r$ filling the remaining budget). \footnote{\small{See project page for additional details, visualizations, and results in other bandwidths or datasets.\label{extend_note}}}. 

\subsection{Transmission performance}
\label{section_performance}
Figs. \ref{JSCC_KODAK_PSNR} and \ref{JSCC_KODAK_SSIM} present PSNR and MS-SSIM performance for image transmission across different channel SNRs.  Implicit-JSCC outperforms all baselines at high SNR, achieving up to $1.89$ dB PSNR and $2.12$ dB MS-SSIM gains, but underperforms at low SNR due to its lightweight design prioritizing low complexity over robustness. Fig.~\ref{JSCC_mp3} reports audio results with $\bm{S^i} \in \mathbb{R}^{32{,}000}$, $\bm{X^i} \in \mathbb{C}^{31{,}750}$, and $6{,}000$ additional channel uses are allocated to $R_\theta$. 
We can observe that Implicit-JSCC consistently outperforms the MP3-Capacity scheme at higher SNRs, validating its modality-agnostic effectiveness.

\subsection{Decoding efficiency and Encoding flexibility}
Implicit-JSCC supports flexible trade-offs between performance and decoding complexity. As shown in Fig. \ref{psnr_complexity_kodak}, it achieves comparable or superior performance with up to $1000\times$ fewer decoding operations and significantly fewer parameters. Fig. \ref{JSCC_KODAK_PSNR_complexity_SNR} additionally shows that low SNRs benefit from larger models, while medium-sized models are more effective at high SNRs due to simpler learning dynamics.
The overfitting paradigm allows flexible encoding strategies. As shown in Table~\ref{tab_Encoding_complexity}, competitive reconstruction can be achieved within $10{,}000$–$20{,}000$ steps ($15$ ms per step), while decoding takes just $2$ ms. Overall, Implicit-JSCC offers strong performance with minimal decoding cost, ideal for scenarios with one encoding and repeated decodings.

\subsection{Visualizations}
Fig.~\ref{fig_Vis_each_module} visualizes the channel input, output, and ReDU output, with color intensity reflecting power allocation for the channel symbols. We can observe that the optimized channel input can encode semantic content robustly and allocate more power to critical textures, focusing on coarse structures (e.g., shapes) at low SNRs and finer details (e.g., digits) at high SNRs. Comparing ReDU output with the channel output reveals that ReDU naturally functions as a denoiser and refiner, suppressing noise at low SNRs and enhancing details at high SNRs, despite not being explicitly trained for this.
\section{Conclusion}
We presented Implicit-JSCC, an overfitted DeepJSCC paradigm that eliminates the need for training datasets and pretrained model storage, reduces decoding complexity by up to $1000$ times, and adapts across modalities through instance-specific optimization. Its efficiency and strong performance make it well-suited for streaming scenarios, where one-time encoding supports repeated decoding, highlighting the potential of instance-level optimization in future DeepJSCC systems.

\bibliographystyle{IEEEbib}
\bibliography{ref}

\end{document}